\documentclass{article}

\usepackage[square,numbers]{natbib}
\usepackage[final]{neurips_2023_ml4ps}

\usepackage[utf8]{inputenc} 
\usepackage[T1]{fontenc}    
\usepackage{hyperref}       
\usepackage{url}            
\usepackage{booktabs}       
\usepackage{amsfonts}       
\usepackage{nicefrac}       
\usepackage{microtype}      
\usepackage{xcolor}         
\usepackage{graphicx}
\usepackage{caption}
\usepackage{subcaption}

\title{Deep Learning Segmentation of Spiral Arms and Bars}

\author{%
  Mike Walmsley\thanks{Second affiliation: Jodrell Bank Centre for Astrophysics, Department of Physics \& Astronomy, University of Manchester, Manchester M13 9PL, UK}\\
  Dunlap Institute for Astronomy and Astrophysics\\
  University of Toronto\\
  Toronto, ON M5S 3H4, Canada\\
  \texttt{m.walmsley@utoronto.ca} \\
  \And
  Ashley Spindler \\
  Centre for Astrophysics Research, Department of Physics, Astronomy and Mathematics \\
  University of Hertfordshire \\
  Hatfield, Hertfordshire, AL10 9AB, UK \\
  \texttt{a.spindler@herts.ac.uk} \\
}

\begin{document}

\maketitle

\begin{abstract}
  We present the first deep learning model for segmenting galactic spiral arms and bars. In a blinded assessment by expert astronomers, our predicted spiral arm masks are preferred over both current automated methods (99\% of evaluations) and our original volunteer labels (79\% of evaluations). Experts rated our spiral arm masks as `mostly good' to `perfect' in 89\% of evaluations. Bar lengths trivially derived from our predicted bar masks are in excellent agreement with a dedicated crowdsourcing project. The pixelwise precision of our masks, previously impossible at scale, will underpin new research into how spiral arms and bars evolve. 
\end{abstract}

\section{Introduction}

Astronomers do not know how spiral arms are made.
There are two long-standing theories: density waves \cite{Lin1987density}
and swing-amplification \cite{Sellwood1988spiral}.
Despite decades of work, it is unclear which theory dominates \cite{Hart2018origins,Yu2020spiral}.
Spiral arms also show different characteristics in different galaxies; they vary in arrangement
(e.g. grand design vs. flocculent),
arm count, and arm tightness (pitch angle).
Precise explanations for these characteristics remain elusive.

A closely-related galaxy structure is galactic bars; linear features in the center of galaxies, often connecting two spiral arms \citep{Buta2011GalaxyMorphology}.
Barred spirals make up about 30\% of disk-type galaxies in the nearby universe \cite{Masters2011}.
Bars alter the structure of a host galaxy by funnelling gas and dust from the disk into the galaxy bulge \citep{Combes1990,Athanassoula2002,Khoperskov2018}.
Bars grow over time; whether long and short bars are distinct subclasses or part of a continuum is unclear \citep{Geron2023}.

Progress in understanding both bars and spiral arms has been limited by the difficulty of measuring their shapes \citep{Yu2020spiral,Masters2021_3d,Hart2016}
Shape measurements to-date have been made by either expert inspection \cite{Kennicutt1982}, rule-based image processing \cite{Davis2014, Lin1987density}, or crowdsourcing \citep{Hoyle2011, Hart2016}.

We develop a deep learning model to jointly identify both spiral arms and bars.

Deep learning for galaxy morphology has focused almost exclusively on classifying entire images (\cite{Dieleman2015,Huertas-Company2015a,Khan2018,Huertas-Company2018,DominguezSanchez2019,Walmsley2020,Walmsley2022decals,Cheng2022,VegaFerrero2021,Walmsley2023desi}, etc.)
Works on segmentation are rare.
\citet{Hausen2020} trained a U-net-like architecture to predict `pixel level morphological classifications' i.e. to identify the pixels belonging to a single galaxy and then ascribe \textit{all} those pixels to a single galaxy class (e.g. spheroid, disk, etc.) 
\citet{Ostdiek2022} used a U-Net-like architecture to classify pixels as belonging to either lensing galaxies or lensed subhalos.
\citet{Richards2023} heavily extended Mask R-CNN to classify both tidal structures around galaxies and background contamination. \citet{Bhambra2022} (hereafter PB+2022) trained a standard CNN (rather than U-Net or Mask R-CNN) to classify galaxy images as barred or not barred, and then used saliency mapping to identify the pixels most relevant to that classification. All of these works adopt a pixelwise classification framework.

\begin{figure}
    \centering
    \includegraphics[trim={0 1.2cm 0 0.5cm},clip]{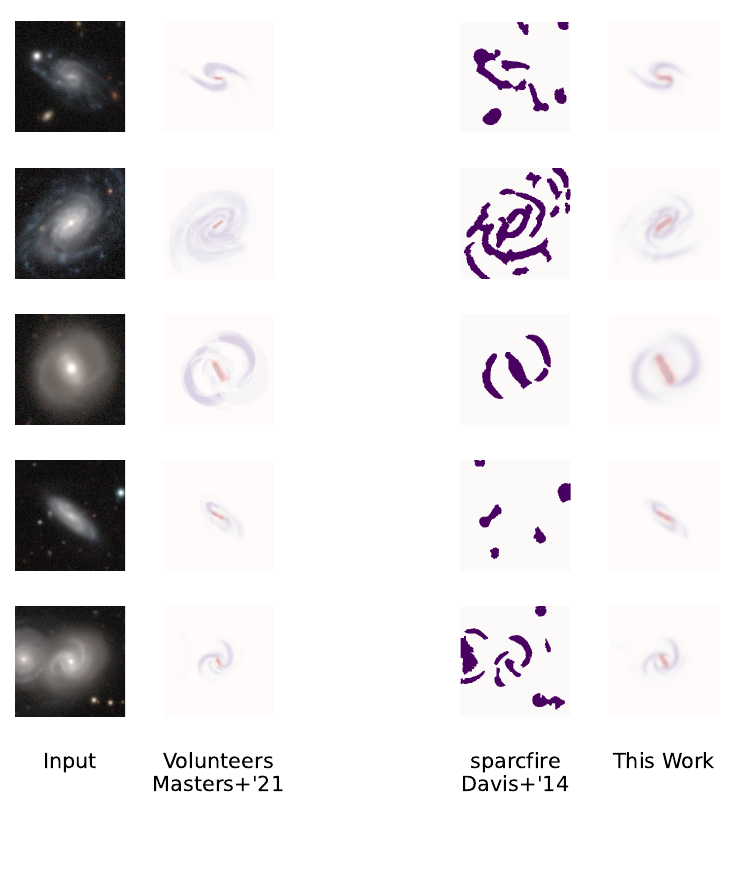}
    \caption{From left: input galaxy image, segmentation masks by volunteers (GZ3D, \cite{Masters2021_3d}), masks by \texttt{sparcfire} \cite{Davis2014}(spiral only), and masks by our model. Rows show randomly-selected test set galaxies, filtered to have GZ DESI \cite{Walmsley2023desi} vote fractions above 0.5.}
    \label{fig:comparison_grid}
\end{figure}

\begin{figure}
    \centering
    \begin{subfigure}[b]{0.58\textwidth}
        \includegraphics[width=\textwidth]{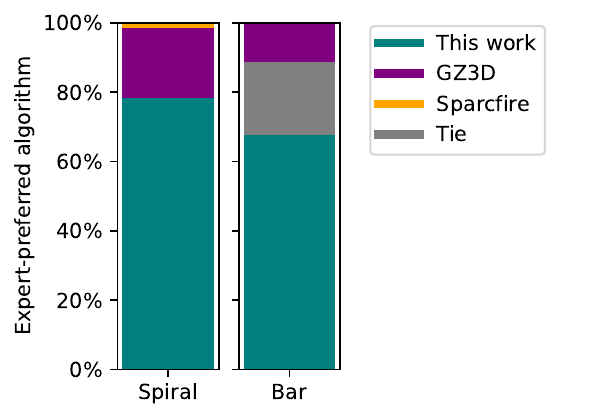}
        \caption{Fraction of astronomer evaluations selecting each `algorithm' as best-performing (N=367, 133). Astronomers strongly prefer our model, even over the original labels we use for training.}
        \label{fig:expertpreferences}
    \end{subfigure}
    \hspace{0.1cm}
    \begin{subfigure}[b]{0.4\textwidth}
        \includegraphics[width=\textwidth]{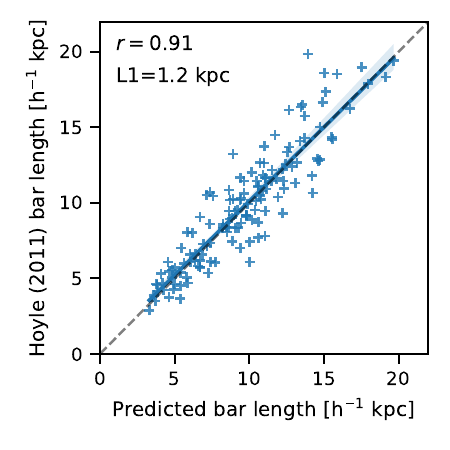}
        \caption{Bar length measurements from our model vs. human measurements (\citet{Hoyle2011}). Length predictions are derived by trivially processing our bar masks.}
        \label{fig:barlengths}
    \end{subfigure}
\end{figure}

\section{Methods}

\subsection{Data and Loss Function Design}

Our segmentation labels are sourced from Galaxy Zoo: 3D, published in \citet{Masters2021_3d}. GZ3D collected crowdsourced pixel masks of the spiral arms and bars of the 29,831 galaxies listed as potential targets for the MaNGA survey \cite{Bundy2015}. While the project was successful and released a public catalogue of pixel masks, \citet{Masters2021_3d} highlighted the time-consuming and labour-intensive nature of the task — particularly in regards to labelling the spiral arms.

GZ3D volunteers used a polygon drawing tool to enclose areas corresponding to bars or spiral arms. Each galaxy was marked by 15 volunteers. We use the individual volunteer polygon vertices published by \citet{Masters2021_3d} to calculate how many volunteers enclosed each pixel.\footnote{Unlike \citet{Masters2021_3d}, we do not discount self-intersecting areas. Many volunteer marks are only self-intersecting where closing line segments slightly intersect, and so discarding these areas removes a significant amount of otherwise-useful annotations.} 

Unlike previous astronomy works, we frame our segmentation task as pixelwise regression. The fraction of volunteers enclosing each pixel reflects the confidence of the crowd. This confidence is a critical training signal when attempting to learn the often-ambiguous task of identifying spiral arms. We choose the loss function:

\begin{equation}
    \mathcal{L}(X,V,w) \propto \:  \sum_{k}^{C} \mid \sum_{i,j}^{X} \sum_{l}^{V} \delta(X_{ij} \in V_{l,k}) - \phi_{w}(X)_{ijk} \mid
\end{equation}
where, for spatial pixel index $(i, j)$ in image $X$, we check if each pixel $X_{ij}$ is enclosed by the vertices marked by volunteer $V_l$ for class $k$ (spiral or bar) and compare that to the model output for the same pixel in channel $k$, $\phi_{w}(X)_{ijk}$. In short, we calculate the mean absolute error when predicting, for each pixel, the fraction of volunteers who included that pixel in their mask, and sum over each output class/channel.

We implement this loss by encoding the pixelwise volunteer vote fractions as greyscale JPEG images (where e.g. a value of 255 corresponds to all 15 volunteers marking a pixel) and then loading and augmenting these masks as if they were conventional images.\footnote{We use only flip and interpolated rotation augmentations, thus preserving the pixel (vote fraction) values.} Our final convolutional layer uses ReLU activation \cite{Agarap2018} to ensure non-negative predictions. The same network jointly learns to predict masks for both spiral arms and bars using a two-channel output.

Our images are sourced from the DESI Legacy Surveys \cite{Dey2018}. DESI-LS images are comparable to the original GZ3D images for identifying spiral arms and bars in nearby galaxies, but using them for training allows for future work to make direct measurements on the full DESI-LS footprint. We use fixed train/validation/test sets with 70/10/20\% division. We only include galaxies with GZ3D masks for spiral arms. We do not penalise the network for predicting bars (or, in principle, spiral arms) on galaxies that volunteers chose not to mark as such, as we judge the lack of marks to be an unreliable indicator for the lack of a feature.

\subsection{Model Architecture}

Our model uses a U-Net architecture which consists of a downsampling encoder and an upsampling decoder. For both submodels, a single step is comprised of two residual blocks, followed by a strided convolution for the downsampling or a transpose convolution for the upsampling. Dropout is used between the residual blocks and convolutions in the encoder to control overfitting. Skip connections between each level of the encoder pass the outputs of the downsampling step to the inputs of the corresponding upsampling step. Following \citet{Smith2022generative} we use Mish activation functions \cite{Misra2019} in the residual blocks. We perform a hyperparameter sweep (on the validation set) to select the batch size, dropout rates, and down/upsampling dimensions. Our code is publicly available at \href{https://github.com/mwalmsley/zoobot-3d}{https://github.com/mwalmsley/zoobot-3d}.

\subsection{Results}

Figure \ref{fig:comparison_grid} presents a random selection of predicted spiral and bar masks. Our predicted masks typically capture spiral and bar features well. Further, they are typically smoother and more precise than the original volunteer labels. The polygon drawing tool used by GZ3D volunteers creates a `blocky' mask which fails to capture the soft edges and ambiguous locations of spiral arms. Small number statistics (each galaxy is marked by 1-15 volunteers) then lead to unphysical step functions where masks from individual volunteers overlap. 
Our model learns from these observed masks to predict the expected mask at each pixel. Unlike the labels, the expected mask is smoothly varying. 

Comparing to previous work is difficult because, to our knowledge, there are no existing automated methods for segmenting spiral arms. The closest comparable code is \texttt{sparcfire} \cite{Davis2014}, which segments as an intermediate step towards mathematically describing the arm segments. 
We run \texttt{sparcfire} with the recommended default parameters, except for disabling deprojection; deprojection transforms the resulting segmentation mask, making direct comparison impossible.
Our model is notably more accurate than sparcfire on the random examples shown. 
We also compared with the recent general purpose `foundation' model SegmentAnything \cite{SegmentAnything2023}; we find that SegmentAnything typically identifies galaxy vs. background pixels but cannot segment structures within galaxies.

To quantitatively assess performance, we recruited 20 astronomers for a blinded comparison between methods. Participants were shown a grid of images (similar to \ref{fig:comparison_grid}) and asked to rate their preferred and second-preferred mask. For each preferred algorithm, they then rated each mask by quality on a four-point scale: `perfect', `mostly perfect (some minor errors)', `mostly poor (some major errors)', and 'totally failed'. Participants made a total of 500 evaluations on 100 galaxies. Participants were not involved in this research and did not know which algorithm generated each mask. 

We find that our model is strongly preferred by astronomers over both \texttt{sparcfire} and our original volunteer labels. Fig. \ref{fig:expertpreferences} shows the fraction of evaluations where astronomers rated each `algorithm' as the best performing; for spiral arms, our model was preferred by an individual expert in 78\% of spiral evaluations (vs. 20\% for the multi-volunteer GZ3D labels and 1\% for \texttt{sparcfire}) and 68\% of bar evaluations (vs. 11\% for the GZ3D labels, with the remainder tied). Our model ranked higher than \texttt{sparcfire} in 99\% of spiral evaluations and higher than GZ3D in 79\% of evaluations \footnote{Formally, the one-tailed Binomial odds of an ensemble of experts preferring our model in $k$ out of $N$ galaxies (excluding ties) under the null hypothesis of no true preference between our model and the original labels is  $p << 0.01$ for spirals ($N=91$, $k=74$) and $p = 0.04$ for bars ($N=26$, $k=17$)}. Astronomers rated our spiral arm masks as ‘mostly good’ to ‘perfect’ in 89\% of evaluations, vs. 77\% for the GZ3D labels.

To quantitatively demonstrate that our masks are useful for astronomy, we process our bar masks to calculate bar lengths. We calculate the length of the confidently-barred\footnote{Defined as having predicted bar vote fractions greater than the median-confidence pixel} pixel mask and convert this length to physical distances. We find (Fig. \ref{fig:barlengths} excellent agreement with the measurements gathered by \citet{Hoyle2011} in a crowdsourcing project dedicated to finding bars.

\section{Conclusion and Discussion}

We have presented the first deep learning model for segmenting spiral arms and bars. Our model outperforms both the closest existing automated approach and the original crowdsourced labels on which it was trained, as judged by independent astronomers in a blinded evaluation.

Accurate segmentation of spiral arms and bars is especially relevant for Integral Field Spectroscopy (IFS) surveys, which measure spectra at every pixel.
This was the original motivation for GZ3D. There are now multiple large IFS surveys (e.g. MaNGA, SAMI) and IFS is available on observational platforms including NIRSpec and MIRI on JWST and ESO-MUSE on the VLT. Accurate automated segmentation masks could enhance research with any of these platforms.

Accurate segmentation is also relevant for the very largest surveys. Our model is trained on DESI-LS images, enabling spiral and bar shape measurements for galaxies with upcoming DESI spectra. Euclid \cite{Laureijs2011} will shortly begin resolving the detailed morphology of higher-redshift ($z \approx 1$) galaxies, allowing us to measure bars and spirals as they form and grow to their present-day appearance.

We chose to use a U-Net architecture because it is appropriate for smaller labelled datasets (here, 5054 training galaxies) and to focus this work on introducing a first solution to this new task. More recent architectures may improve performance.

In closing, we would like to note a negative result. We found that, by adding a Dirichlet regression head \cite{Walmsley2023zoobot} to the latent space of our U-Net, we could also predict Galaxy Zoo volunteer vote fractions competitively with a dedicated classification model. However, adding this joint prediction task did not improve the quality of our segmentation masks. This complicates the `astronomy foundation model' view presented by \citet{Walmsley2022Towards} (see also \cite{Rozanski2023,Leung2023,Slijepccevic2023}). General purpose models may not always outperform dedicated ones.

\begin{ack}
We are grateful to the Zooniverse volunteers and GZ:3D authors for creating the labels we rely on here.

We would like to thank the astronomers who contributed blinded assessments, including: Dominic Adams, Elisabeth M Baeten, Hugh Dickinson, James Tropp Garland, Antonio Herrera-Martin, Bill Keel, Karen L Masters, James Pearson, Jonathon C S Pierce, Shravya Shenoy, Brooke Simmons, Josh Speagle, Grant Stevens, Laura Trouille, and Klaas Wiersema.

This work was supported by a grant from Meta.
MW is a Dunlap Fellow. The Dunlap Institute is funded through an endowment established by the David Dunlap family and the University of Toronto.
This work has made use of the University of Hertfordshire high-performance computing facility (\url{https://uhhpc.herts.ac.uk/}.

\end{ack}

\bibliography{references}

\end{document}